\documentstyle[12pt]{article}

\def\sqr#1#2{{\vcenter{\hrule height.#2pt
   \hbox{\vrule width.#2pt height#1pt \kern#1pt
    \vrule width.#2pt} 
    \hrule height.#2pt}}}

\newcommand{\be}{\begin{equation}}
\newcommand{\ee}{\end{equation}}
\newcommand{\beq}{\begin{eqnarray}}
\newcommand{\eeq}{\end{eqnarray}}

\begin{document}
\begin{titlepage}
\begin{flushright}
UFIFT-HEP-99-2
\end{flushright}
\vspace{.4cm}
\begin{center}
{\large \bf Energy Density and Pressure of Long Wavelength Gravitational
Waves}
\end{center}

\begin{center}
L. R. Abramo$^{* \dagger}$
\end{center}
\begin{center}
\textit{Department of Physics \\ University of Florida \\
Gainesville, FL 32611 USA}
\end{center}


\begin{center}
ABSTRACT
\end{center}

\noindent \hskip .5cm Inflation leads us to expect a spectrum of
gravitational waves (tensor perturbations) extending to wavelengths much
bigger than the present observable horizon. Although these gravity waves
are not directly observable, the energy density that they contribute
grows in importance during the radiation- and dust-dominated ages of the
universe. We show that the back reaction of tensor perturbations during
matter domination is limited from above, since gravitational waves of
wavelength $\lambda$ have a share of the total energy density $\Delta
\rho(\lambda)/\rho$ during matter domination that is at most equal to the
share of the total energy density that they had when the mode $\lambda$
exited the Hubble radius $H^{-1}$ during inflation. This work is to be
contrasted to that of Sahni\cite{Sahni}, who
studied the energy density of gravity waves only insofar as their
wavelengths are
smaller than $H^{-1}$. Such a cut-off in the spectral energy of gravity
waves leads to the breakdown of energy conservation, and we show that this
anomaly is eliminated simply by taking into account the energy density and
pressure of long wavelength gravitational waves as well as short
wavelength ones.

\begin{flushleft}
\end{flushleft}
\vspace{0.5cm}
\begin{flushleft}
$^*$ Address after June 1, 1999: Theoretische Physik, 
Ludwig Maximilians \\
Universit\"{a}t, Theresienstr. 37, D-80333 M\"{U}NCHEN, GERMANY \\
$^{\dagger}$ e-mail: abramo@phys.ufl.edu \\
\end{flushleft}
\end{titlepage}


\section{Introduction}

Gravitational waves are quantum-mechanically produced in all inflationary
scenarios \cite{GW_Grish,GW_Starob,GW_Inf1,GW_Inf2}. In some models
gravitational waves are a crucial component of the cosmic microwave
background\cite{GW_CMB}, and the prospects that gravitational waves
generated during inflation can be directly observed in the future using
space-based interferometry are not beyond hope\cite{GW_LIGO}.

Scales that are now entering our Hubble radius ($H_0^{-1} \approx
10^{28}$cm)
correspond to physical wavelengths that were equal to the Hubble radius
around 65 e-foldings of the scale factor before the end of inflation.
Gravity waves that were generated earlier than that time during inflation
are today larger than the present Hubble radius and vice-versa.
Therefore,
unless we fine-tune inflation to happen for no more than about 65
e-foldings, there must be gravitational waves (as well as density
perturbations) with physical wavelengths much bigger than $10^{28}$cm. 

Obviously, these long wavelength gravitational waves cannot be directly
detected, nor have they any impact on observables such as the CMB. Even
if the spectrum of gravitational waves on sub-horizon scales is measured,
we still could only guess what the spectrum might look like for the
super-horizon modes.

Nevertheless, cosmological perturbations of long wavelengths can have an
impact on the background in which they propagate, through their
self-energy and their gravitational interactions. Tsamis and
Woodard\cite{TW}, for example, have investigated the feedback from
quantum mechanical pair production in pure gravity with a cosmological
constant, and found that the two loop back reaction of the metric
fluctuations have the effect of screening the cosmological constant. The
one loop back reaction of fluctuations during scalar-field inflation has
also been studied by the present author and collaborators, and in this
case we found that in some models of inflation the expansion rate of the
universe slows down faster due to these feedback effects\cite{MAB,ABM,AW}. 

There is a very simple physical picture for these processes\cite{TW}: 
during inflation, virtual pairs are created all the time, and eventually
some of them become trapped in the expansion of the universe. As the pair
is pulled apart by inflation, the gravitational potential that must exist
between the pair fills the intervening space. Even after the pair becomes
causally disconnected, the gravitational potential still remains, just as
the potential of a particle that falls into a black hole remains after
the particle crosses the hole's horizon. Since these gravitational
interactions are attractive, we expect them to have a tiny impact in
slowing the expansion of the universe, as they try to bring the pair back
together. The only questions are what is the strength of back reaction,
and what is its time dependence (whether back reaction effects grow or
decrease in time, and if they can ever become important). 

In this paper we examine the energy density and pressure engendered by
gravitational waves at lowest order in perturbation theory (one loop).
This effective energy-momentum tensor of gravitational waves, when plugged
into Einstein's field equations, is the source for the back reaction of
the gravitational waves on the expansion rate of the homogeneous and
isotropic universe. We analyze these effective terms during and after
inflation, focusing on the contribution from long wavelength
modes\footnote{Short wavelength gravity waves have been extensively dealt
with in the standard textbooks\cite{Weinberg}. The relevant fact in that
limit is that the kinetic energy of short wavelength gravitons is much
more important than their gravitational interactions, and the gravitons
behave essentially like conformally invariant ultra-relativistic particles
- i.e., their energy density falls like radiation, $a^{-4}(t)$, where
$a(t)$ is the scale factor.}. Sahni\cite{Sahni} considered the energy
density due to a spectrum of gravitational waves extending to wavelengths
much bigger than the Hubble radius, but in that work a gravitational wave
mode is taken into account only if and when the mode becomes smaller than
the Hubble radius $H^{-1}(t)$ -- an idea first proposed, at least in the
context of gravitational waves from inflation, by Allen
\cite{GW_Inf2}. We show that this cut-off leads to the breakdown of
energy-momentum conservation, or, equivalently, to a violation of the
Bianchi identities. The most natural, and the simplest, way of avoiding
this unconventional feature is to include the energy density and pressure
of gravitational waves when their wavelengths are bigger than the Hubble
radius.

The backgrounds considered for this work are flat
Friedmann-Robertson-Walker space-times where the scale factor is a
power-law of time, $a(t) \propto t^s$ with $s>0$. However, we allow the
value of $s$ to change during the evolution of the universe: we assume
$s>1$ during inflation ($t<0$) and $s<1$ afterwards ($t>0$). 

This paper is organized as follows: in section 2 we determine the
perturbative background, then solve (exactly) the equations for the
gravitational waves. In section 3 we discuss the nonlinear terms that
give rise to back reaction and that were discarded in section 2, and show
how gravity waves can impact the background. We also show that the back
reaction of long wavelength gravitational waves is not only consistent
with, but in fact is demanded by conservation of energy and momentum. We
employ the resulting formulas in section 4 for a generic
power-law inflationary
universe that ``reheats" at $t=0$, and show that the share of the total
energy due to long wavelength modes grows during the
decelerated expansion phase ($s<1$), but is limited from above. We
conclude in section 5.


\section{Power-law backgrounds}

Consider a flat Friedmann-Robertson-Walker metric,

\be
\label{FRWmetric}
ds^2 = -dt^2 + a^2 (t) d\vec{x} \cdot d\vec{x} \; ,
\ee
where the scale factor is a power-law of time,

\be
\label{a_t}
a(t) = \left( 1 + \frac{H_i t}{s} \right)^s \quad \quad s>0 \; .
\ee
The expansion rate is given by the Hubble parameter,

\be
\label{H_t}
H(t) = H_i \left (1+ \frac{H_i t}{s} \right)^{-1}
\ee
where $H_i= H (t=0)$. From the Einstein equations for the background

\beq
\label{EFE_0}
3H^2 = \frac{\kappa^2}{2} \rho (t) \; , \\ 
\label{EFE_i}
-3H^2 - 2 \dot{H} = \frac{\kappa^2}{2} p (t) \; ,
\eeq
where $\kappa^2 = 16 \pi G$ and a dot indicates a time derivative,
it follows that

\be
\label{rho_t}
\rho (t) \propto \left( 1 + \frac{H_i t}{s} \right) ^{-2} 
\quad \quad , \quad \quad 
\frac{p}{\rho} = -1 + \frac{2}{3s} \; .
\ee
Notice that the de Sitter limit $s \rightarrow \infty$ is well defined in
this time parameterization.

When we include gravitational waves the metric reads\cite{revpaper}

\be
\label{metric}
ds^2 = -dt^2 + a^2 (t) 
	\left[ \delta_{ij} + h_{ij}(\vec{x},t)
	\right] dx^i dx^j \; .
\ee
The gravitational waves $h_{ij} (\vec{x},t)$ (also known as tensor
perturbations) are typically expanded in modes as follows:

\be
\label{h_k}
h_{ij} (\vec{x},t) = \sum_k \left[ 
	\epsilon_{ij}(\vec{k}) h_k(t) e^{-i \vec{k} \cdot \vec{x} } 
	+ {\rm c.c.} \right] \; ,
\ee
where $\epsilon_{ij}$ is the polarization tensor. If one's interest is in
quantizing this system, all that is needed is to promote the fields
$h_k(t)$ to quantum operators and to impose the usual canonical
commutation relations. For now we are only interested in the time
dependence of modes of a given wavenumber. Later we will consider
their amplitudes, which arise due to the well-known mechanism of quantum
mechanical pair creation in an expanding universe, also
known as superadiabatic amplification\cite{GW_Grish}.

The equation obeyed by the modes $h_k$ is obtained through the
linearization of Einstein's field equations, and is found to be identical
to that obeyed by a minimally coupled, massless scalar: 

\be
\label{eq_h}
\ddot{h}_k + 3H \dot{h}_k + \frac{k^2}{a^2(t)} h_k = 0 \; .
\ee
This equation can be exactly solved in power-law backgrounds 
(see, e.g., \cite{Sahni}). For the sake of clarity we briefly re-derive
these solutions in what follows.

It is useful to introduce the variable

\be
\label{y}
y(k,t) = k\eta (t) \equiv k \int^t \frac{dt'}{a(t')} = 
\left\{ 
\begin{array}{ll}
\frac{k}{H_i} \frac{s}{1-s} \left( 1+ \frac{H_i t}{s} \right)^{1-s} 
& s \neq 1 \; , \\
\frac{k}{H_i} \log{ \left( 1+ \frac{H_i t}{s} \right) }
& s = 1 \; ,
\end{array}
\right.
\ee
where $\eta$ is the conformal time. Notice also that the far
infrared is given by the limit $|y| \rightarrow 0$. Indeed, for $s\neq 1$
we have $y(k,t)=\frac{k}{a(t)H(t)} \frac{s}{1-s}$, that is, a mode is
infrared if its physical wavelength is bigger than the Hubble radius at
time $t$.

If we now write the modes $h_k$ in the form

\be
\label{h_F}
h_k = y^\nu F_\nu (y) \quad \quad \quad \nu(s) 
= \frac12 \frac{3s-1}{s-1} \; ,
\ee
then the equation for $F_\nu (y)$ can be reduced to the form

\be
\label{eq_F}
y^2 F'' + y F' + (y^2 - \nu^2) F = 0 \; ,
\ee
which we recognize as Bessel's equation. Thus, gravitational wave
modes in power-law backgrounds are written as

\beq
\label{h_JY}
h_k(y) & = & y^\nu 
	\left[ A_k H_{|\nu|}^{(1)} (|y|) + B_k H_{|\nu|}^{(2)} (|y|) 
	\right] \; , \\
\label{h_JJ}
	& = & y^\nu 
	\left[ \tilde{A}_k J_{|\nu|} (|y|) + \tilde{B}_k J_{-|\nu|} (|y|) 
	\right]  \; , \quad \quad \quad |\nu| \neq {\rm integer} \; , 
\eeq
where $H_{|\nu|}^{(1,2)}$ are Hankel functions of the first
and second kind. The positive energy eigenmodes of the gravitational
waves are associated with the Hankel functions of the second kind.
In going from the first to the second line we used the property that for
non-integer $|\nu|$ the Bessel functions $J_{|\nu|}$ and $J_{-|\nu|}$ are
linearly independent.

In the ultraviolet limit $|y| \rightarrow 
\infty$ the dominant behavior of the Bessel functions is an
oscillating term $\exp{[-i y(t)]}/a(t)$. Therefore, in that limit
gravitational waves can be regarded as simple plane waves.

The infrared limit is far richer. $J_{\nu}(|y|)$ has the following asymptotic 
expansion when $|y| \rightarrow 0$:

\be
\label{J_as}
J_{\nu} \approx \left( \frac{|y|}{2} \right)^{\nu}
	\frac{1}{\Gamma(\nu+1)} 
	\left[ 1 - \frac14 \frac{1}{1+\nu} |y|^2
	+ {\cal O}\left({|y|^4}\right) \right] \; , 
\ee
and thus from (\ref{h_F}) and (\ref{h_JJ}) we have

\beq
\label{h_as}
h_{k} (t) &\approx& 
	C_k |y|^{2\nu} \left[ 1 - \frac{1}{4(1+\nu)} |y|^2 + 
	{\cal O}(|y|^4) \right] \\
\nonumber
&&+ 	D_k |y|^{0} \left[ 1 - \frac{1}{4(1-\nu)} |y|^2 +
	{\cal O}(|y|^4) \right] \; . 
\eeq

If $y<0$ ($s>1$, $\nu>3/2$) then $y \rightarrow 0^-$ in time.
From Eq. (\ref{h_as}) we have that when $s>1$, $|y|$
becomes smaller with time and therefore the term proportional to $C_k$ is
subdominant.

If, on the other hand, $y>0$ ($s<1$, $\nu<0.5$) then $y$ grows as a
function of time. However, if $1/3<s<1$ then $\nu<0$ and the term after
$C_k$ is subdominant again. Lastly, if $0<s<1/3$ then $y$ is growing in
time, therefore the first term in (\ref{h_as}) looks like it will become
dominant with respect to the term after $D_k$.

At this point we should remind the reader of the cosmological scenario we
are considering: initially there is an inflationary phase $s_1>1$ during
which tensor perturbations (gravity waves) are produced and stretched
towards the infrared ($|y|$ decays with time when $s>1$). Later, when the
universe relaxes to a phase $s_2<1$, gravity waves become less and less
infrared ($|y|$ now grows with time), with some waves eventually
re-entering the Hubble radius and becoming effectively ultraviolet.

By the time of the transition from the inflationary phase to the ``normal
matter" phase (which we can fix at $t=0$), the only pieces of the
gravitational waves that survived from the inflationary phase were the
dominant modes $D^{(1)}_k$. These modes must be glued on the $t=0$
space-like hypersurface to the gravitational wave solutions of the phase
$s_2<1$. By matching the solutions and their first time derivatives
on the $t=0$ surface, we get the following expressions for the amplitudes
in each mode in the $s_2<1$ phase:

\beq
\label{ampl_D}
D^{(2)}_k &=& D^{(1)}_k \left[ 1 + 
	{\cal O} \left( \frac{k}{H_i} \right)^{2} \right] \; , \\
\label{ampl_C}
C^{(2)}_k &=& D^{(1)}_k
	\left( \frac{s_1}{s_1+1} -\frac{s_2}{s_2+1} \right)
	\frac{s_2}{3s_2-1}
	\left( \frac{1-s_2}{s_2} \right)^{2\nu_2} \\
\nonumber && \times
	\left( \frac{k}{H_i} \right)^{2-2\nu_2} 
	\left[ 1 + {\cal O} \left( \frac{k}{H_i} \right)^{2} \right] \; , 
\eeq
where $\nu_2\equiv \nu(s_2)$. We see then that for $1/3 <s_2<1$ the
dominant mode of the inflationary phase $s_1>1$ is completely transmitted
to the dominant mode of the $s_2<1$ phase. For $s<1/3$ the situation is
slightly more complex: the initial amplitude of the growing mode is
tiny when compared to the amplitude of the decaying mode. The growing
mode only surpasses the decaying mode when $k/aH \gg 1$, but
by then the wavelength of the gravitational wave is already much
smaller than $H^{-1}$, and the infrared limit is not valid anymore.

We end this section with the general result for the dominant modes in the
asymptotic expansion for the gravitational waves in the infrared limit:

\be
\label{h_asympt}
h_k(t) \approx D_k \left[ 1 - \frac12 \frac{s^2}{1-s^2} 
	\left( \frac{k}{aH} \right)^2 
	+ \ldots \right] \; ,
\ee
where we have substituted $y = \frac{s}{1-s}\frac{k}{aH}$ into
Eq. (\ref{h_as}). Although the expression above was derived for
non-integer $|\nu|$, one can show that
(\ref{h_asympt}) follows if one takes $|\nu|$ integer as well.


\section{The Back Reaction of Gravitational Waves}

Gravitational waves have an impact on the background in which they
propagate. This nonlinear feedback, called back reaction, has been
discussed extensively in the case where the gravitational waves are in
the so-called ``geometrical optics" limit, that is, when their
wavelengths are much smaller than the curvature radius of the background
space-time in which they travel\cite{Weinberg,Isaacson}. In this limit
(the ultraviolet, by our definitions of the last section) the energy
density of a gravitational wave mode $h_k^{UV} = c_k \exp{[iy(t)]} / a(t)
$ is purely kinetical, and is given by the pieces of the $0-0$ component
of the Einstein tensor which are quadratic in the amplitude of the metric
fluctuations\cite{Weinberg}: 

\beq
\Delta \rho^{UV} &\equiv& \frac{2}{\kappa^2} G_{00}^{(2)} 
	\approx \frac{1}{4\kappa^2} \left( |\dot{h}_k| + 
	|\vec{\nabla} h_k \cdot \vec{\nabla} h_k|
	\right) \\
\label{Delta_rho_UV}
	&=& \frac{1}{2\kappa^2} \frac{k^2}{a^4(t)} |c_k|^2 \; ,
\eeq
that is, the energy density of high frequency gravity waves falls off
like the energy density of radiation.

The limit where gravitational waves have long wavelengths is much less
discussed in the literature, chiefly, in our opinion, due to the
misguided belief that long wavelength fluctuations could not have any
impact on the expansion of the universe as measured by a ``local"
observer.

To gain insight into this question we propose the following thought
experiment: consider an inertial observer during inflation that throws
his general relativity textbook away, and starts measuring the
gravitational potential of the book. Since inflation takes small
coordinate distances into enormous physical distances, soon this observer
watches as his book falls out of his Hubble radius $H^{-1}$. The question
is: when the book falls out of contact with the observer (that is,
when the physical distance between observer and book exceeds $H^{-1}$),
what happens to the gravitational potential of the book: does it vanish
completely, or is there some small potential remaining? We believe that the
only physically acceptable answer, and in particular the only answer 
consistent with conservation of energy and momentum, is that the
observer still measures the gravitational pull from the book.

The obvious analog of this thought experiment is a similar observer that
throws his book into a black hole of mass $M$, and measures the
gravitational pull of the book with mass $m$. In this case the answer to
the question ``What happens to the gravitational potential of the book
after it falls through the horizon?" is clear: the observer still feels
the pull of the book's gravity, since now the black hole has a mass
$\tilde{M}=M+m$. The standard explanation is that the persistent
gravitational potential of the book is due to virtual gravitons that were
emitted near the horizon just as the book fell through the
horizon into the black hole.

Notice that both observers are ``local", that is, they have no knowledge
of what goes on beyond either the cosmological or the black hole
horizons. Nevertheless, both are able to measure the build-up of the
book's gravitational potentials, even after the book has lost causal
contact with the observers. At a much later time neither observer will
be able to tell the difference between the pull of the book and that due
to the background accelerations. Indeed, all that the observers {\it can}
measure at that point are accelerations - acceleration towards the black
hole in the latter example, cosmic accelerations in the former example.
Therefore, neither observer ``sees"  the book anymore, although they
certainly feel the effects of the book's gravitational pull. 

By the same token, long wavelength perturbations can have a gravitational
impact on the accelerations (i.e., expansion rate) of the background
space-time. The physical picture is as follows: perturbations are
generated causally, inside the Hubble radius, and as they are redshifted
by inflation their gravitational interactions fill the intervening space. 

Rather than ``crossing" the Hubble radius, the correct statement is to
say that for an inertial observer these fluctuations become exponentially
frozen at the Hubble radius as their physical wavelengths become larger
than $H^{-1}$, much like the book that falls into the black hole appears
to the inertial observer to be frozen near the black hole horizon. The
(by now long wavelength) perturbations remain frozen near the Hubble
radius for the duration of inflation, and only after reheating they start
to defrost (since after inflation $H^{-1}$ grows faster than physical
wavelengths). Eventually, as the Hubble radius grows and we have access
to larger and larger distances, the perturbations becomes accessible to
local observers who can then detect them directly. 

At no point in time the inflationary perturbations ever fall completely
out of contact with the local observer that witnessed their migration
from small-scale fluctuations, to long wavelength perturbations, to
cosmological perturbations. This is the crucial distinction between a
cosmological scenario where perturbations are generated by a causal
process (inflation), and the ``old"  scenario of the radiation- then
dust-dominated ages of the universe where the spectrum of perturbations
had to be imposed by {\it fiat}: in the ``old" scenario there is a true
particle horizon in both ages ($R_H = t/3$ in the radiation phase),
whereas in the inflationary scenario the Hubble radius is only an
apparent particle horizon -- the real particle horizon is the physical
size of the quasi-homogeneous region that expanded coherently from the
beginning of inflation, and is usually many orders of magnitude larger
than the apparent horizon.

It appears therefore that there should be persistent gravitational
interactions engendered by inflationary perturbations, regardless of the
wavelength of those perturbations. The only questions are what is the
magnitude of this effect, and whether it becomes more or less important
in time.

The simplest way of estimating the importance of gravitational back
reaction is by computing the lowest-order nonlinear (quadratic) 
corrections to the Einstein field equations, and comparing them to the
background energy density and pressure. If we had used quantum mechanics
and perturbation theory consistently, and if we also included the
fluctuations in the matter fields that drive inflation, this calculation
would correspond to computing the one loop effective theory (in that
respect see \cite{AW} and \cite{ITTW}).

In what follows we calculate the one loop energy-momentum tensor for
gravity waves, but do not solve for its back reaction on the metric and
the expansion rate of the universe. The distinction should be clear: the
former is a source term, while the latter are the actual solutions of the
Einstein equations where the quadratic corrections have been taken into
account. The reason we avoid writing down and solving these simple
equations is purely economical: the effective energy-momentum tensor for
gravity waves, at least within the scope of second order perturbation
theory, never becomes important when compared to the energy-momentum
tensor of the background.

The Einstein field equations to second order give the following
expressions for the energy density and pressure of gravitational
waves\cite{ABM}: 

\beq
\label{rho_GW}
\frac{\kappa^2}{2} \Delta \rho &\equiv& G_{00}^{(2)} = \left[
	H \dot{h}_{ij} h_{ij} + 
	\frac18 \left( \dot{h}^2_{ij} + \frac{h_{ij,k}^2}{a^2}
	\right) \right]\; , \\
\label{p_GW}
\frac{\kappa^2}{2}
\Delta p &\equiv& \frac{G_{ii}^{(2)}}{3} =
	\frac{1}{24} \left( 
	-5 \dot{h}^2_{ij} + 7 \frac{h_{ij,k}^2}{a^2}
	\right) \; , 
\eeq
where the Latin indices are summed with the Euclidean metric. 
The expression for the pressure was obtained through the 
assumption (valid at least for inflation-generated tensor
fluctuations) that the spectrum of gravity waves does not break
the homogeneity and isotropy of the background space-time.

In the ultraviolet limit the first term in the right-hand-side of Eq. 
(\ref{rho_GW}) can be neglected, and the familiar result
(\ref{Delta_rho_UV}) follows. It is easy also to calculate the pressure
of gravitational waves in this limit and obtain the expected equation of
state, $\Delta p^{UV} / \Delta \rho^{UV} = 1/3$.

The drag term $H\dot{h} h$ is an additional interaction of the
gravitational waves that is only important for super-horizon waves. Even
though $H\dot{h}h <0$, it seems difficult for us to interpret it
as a gravitational potential term, since the Newtonian potential
does not appear at this order in perturbation theory.

Eqs. (\ref{rho_GW})-(\ref{p_GW}) should be consistent with conservation
of energy and with the Bianchi identities (which are one and the same
thing here). Since we are including quadratic terms in Einstein's
equations, we expect the Bianchi identities to take the usual form with
maybe some quadratic corrections. Indeed, if we truncate the perturbative
expansion of the Bianchi identities to quadratic order in the
gravitational waves, we obtain

\beq
\label{Bianchi}
0 &=& \left[ G^\mu_{\; \nu ; \mu} \right]^{(2)} \\
\nonumber
&=&	G^{\mu \; (2)}_{\; \nu , \mu}
	- \Gamma^{\alpha \; (2)}_{\mu\nu} G^{\mu \; (0)}_{\; \alpha}
	- \Gamma^{\alpha \; (0)}_{\mu\nu} G^{\mu \; (2)}_{\; \alpha} 
	+ \Gamma^{\alpha \; (2)}_{\mu\alpha} G^{\mu \; (0)}_{\; \nu}
	+ \Gamma^{\alpha \; (0)}_{\mu\alpha} G^{\mu \; (2)}_{\; \nu} 
	\; , 
\eeq
where we have canceled some terms using that $G^{\mu \; (1)}_{\; \nu} [h]
= 0$ for gravitational waves.

The $0-0$ component of this algebraic identity reads

\be
\label{cons_eq_0}
\frac{d}{dt} \Delta \rho + 3 H (\Delta \rho + \Delta p) + 
	\frac12 \dot{h}_{ij} h_{ij} (\rho + p) = 0 \;.
\ee
It is a short exercise to show that by substituting definitions
(\ref{rho_GW})-(\ref{p_GW}) into (\ref{cons_eq_0}) takes us back to
the equation of motion for the gravity waves, Eq. (\ref{eq_h}).

In the effective Einstein field equations we must add the energy density
$\Delta \rho$ and pressure $\Delta p$ of gravitational waves to the
energy density $\rho$ and pressure $p$ of the background matter. However,
since $\Delta \rho$ and $\Delta p$ obey the modified energy conservation
law (\ref{cons_eq_0}), it is useful to define the {\it effective}
pressure\cite{ABM}

\be
\label{p_eff}
\frac{\kappa^2}{2} \Delta p_{\rm eff} \equiv 
\frac{\kappa^2}{2} \Delta p - \frac{\dot{H}}{3H} \dot{h}_{ij} h_{ij} \; ,
\ee
where we have used the background relations (\ref{EFE_0})-(\ref{EFE_i}) 
to simplify the expression. The last term in (\ref{p_eff}) is due to the
appearance of the additional interaction $H\dot{h}{h}$ in (\ref{p_GW}),
and is only important for long wavelength gravitational waves. In terms
of this effective pressure, the equation of conservation of energy for
the gravity waves takes its usual form,

\be
\label{cons_eq}
\Delta \dot{\rho} + 3 H (\Delta \rho + \Delta p_{\rm eff})
	= 0 \; .
\ee
Eq. (\ref{cons_eq}) is an algebraic constraint on the time dependence of
the energy density of gravitational waves. In particular, this constraint
is valid even by the time when the physical scales of the gravitational
waves are becoming larger than the Hubble radius.

If one believes that
the energy density and pressure of gravitational waves effectively
disappear after they cross the Hubble radius (i.e., that $\Delta \rho$
and $\Delta p_{\rm eff}$ are exponentially suppressed, rather than
power-law suppressed), then one should at least show that at that point
the energy density and effective pressure cancel each other in Eq.
(\ref{cons_eq}),
$\Delta p_{\rm eff} \approx - \Delta \rho$. The same is true if one
believes that a gravitational wave only acquires energy upon entering the
Hubble radius. Otherwise one is forced to the heterodox conclusion that
energy is not conserved, $\Delta \dot\rho + 3 H (\Delta \rho + \Delta
p_{\rm eff}) \neq 0$, and to some modification of Einstein's equations
that includes sources and sinks of particles in order to account for the
matter creation entailed by the non-conservation of energy\cite{Maia}.

Furthermore, because the Bianchi identities are integrability conditions
on the classical equations of motion in curved space-time, it is
difficult for us to understand how one could assign arbitrary energy
density and pressure to the super-horizon gravitational waves while still
keeping the equations of motion that those waves should obey unchanged.

For example, in Ref. \cite{Sahni} it is easy to see that energy
conservation is violated: the integrated energy of short wavelength modes
is given by Eq. (11) of that paper, which is just a sum over momentum
modes of the energy per mode given in our expression
(\ref{Delta_rho_UV}). Since that author chose to neglect modes of
physical wavelengths larger than the Hubble radius, he picked an infrared
cut-off corresponding to the time-dependent comoving scale $k_0(t)=H(t)$.
It is clear that such a time-dependent comoving cut-off injects an extra
time dependence into the integrated energy density and violates energy
conservation (for ultraviolet modes $\Delta p \approx \Delta p_{\rm
eff}$, so even the naive energy conservation law is violated). What is
happening, of course, is that in \cite{Sahni} each mode that comes inside
the Hubble radius suddenly starts to contribute to the energy density,
that is, for energy accounting purposes that mode is effectively
``created"  at the time when $\lambda_{\rm phys} \approx H^{-1}$. By the
same token, these modes were effectively ``destroyed" when they left the
Hubble radius during inflation. 

The simplest way to eliminate this puzzling anomaly is to include in the
calculations the energy density of gravity waves regardless of their
wavelength. If one does that then what we have shown is that energy is
conserved as it should, and there is no need to invoke matter creation
or any other non-standard physics.

Up to this point the discussion in this section has been generic, and
applies to gravitational waves in any background. Now we compute the
energy density and pressure contributed by long wavelength gravitational
waves in the power-law backgrounds $a = (1+ H_i t/s)^s$. From Eq.
(\ref{h_asympt}) we have that

\be
\label{hdot}
\dot{h}_k = - D_k (k) \frac{s}{1+s} H 
	\left[ \left( \frac{k}{aH} \right)^2 
	+ {\cal O} \left( \frac{k}{aH} \right)^4 \right] 
\; .
\ee
After substituting Eq. (\ref{hdot}) into expressions (\ref{rho_GW}),
(\ref{p_GW}) and (\ref{p_eff}) and keeping the leading terms in $k/aH$
one obtains for the energy density and pressure of the infrared gravity
wave modes:

\beq
\label{rho_LW_s1}
\frac{\kappa^2}{2} \Delta \rho (k) &=& 
	- \frac{|D_k(k)|^2}{8} \frac{7s-1}{s+1} \frac{k^2}{a^2} 
	+ \ldots \; , \\
\label{p_LW_s1}
\frac{\kappa^2}{2} \Delta p_{\rm eff} (k) &=& 
	\frac{|D_k(k)|^2}{24} \frac{7s-1}{s+1} \frac{k^2}{a^2} 
	+ \ldots \; ,
\eeq
that is, $\Delta p_{\rm eff} = - \Delta \rho/3 $.
It can be shown that this result holds for the case $s=1$ as well.

We remind the reader that $p=-\rho/3$ is the equation of state of
curvature, so back reaction of long wavelength gravitational waves can be
thought of as a curvature term in addition to the background energy
density.  We expect then that $\Delta \rho \propto a^{-2}$, as indeed is
the case. Notice that $\Delta \rho <0$, that is, the back reaction of
long wavelength gravitational waves tends to slow the expansion rate of
the universe (like a positive curvature would slow the expansion). This
is so because the (negative) drag term $H\dot{h}h$ in (\ref{rho_GW}) is
more important than the (positive) kinetic terms for infrared modes
(actually, the spatial gradient and the drag term are of the same order
of magnitude, but the drag term has a bigger numerical factor). 

Of course, if $s>1$ the energy density of the background falls like
$a^{-2/s}$, thus back reaction never becomes important during inflation.
If $s<1$, on the other hand, $\Delta \rho$ falls {\it less} fast than
$\rho$, and with time the long wavelength gravity waves increase their
share of total energy. Of course, if $s=1$ [which implies $a(t)H(t)=$
constant] then gravitational waves contribute a constant share of the
total energy density, a fact that led to the conjecture\cite{Sahni} that
the universe would tend to an equilibrium phase with $s=1$ 
if the back reaction of gravitational waves ever became important.

In the next
section we show that the growth of the share of energy density
contributed by long wavelength gravitational waves during matter
domination is elusive, since
this share peaks by the time when the gravitational waves come
back inside the Hubble radius. We will show that the maximal value of the
fraction of the total energy density that is contributed by a
gravitational wave in the $s<1$ phase is the fraction of the total energy
density contributed by that wave at the instant when it crossed the
Hubble radius during inflation.


\section{Back Reaction Before and After Reheating}

In this section we study a model in which the universe inflates ($s_1>1$)
when $t<0$, then reheats at $t=0$ and finally expands at a decelerating
rate ($s_2<1$) for $t>0$. The scale factor can be conveniently
parameterized as

\be
\label{a_12}
a(t) = \left\{
\begin{array}{ll} 
	\left( 1 + \frac{H_i t}{s_1} \right)^{s_1} 
	& t \leq 0 \\
	\left( 1 + \frac{H_i t}{s_2} \right)^{s_2} 
	& t \geq 0 \; ,
\end{array} \right.
\ee
so that $a(t=0)=1$ and $H(t=0)=H_i$.

Consider now the physical wavelength $\lambda_1^p$ that crosses the
Hubble radius at $t=t_1<0$, that is, the scale of the comoving momentum
$k_1 \equiv 2\pi a(t_1)/\lambda^p_1 = H(t_1) a(t_1)$. This scale will
cross back into the Hubble radius at some time $t_2 >0$ given by the
solution of

\be
\label{t_12}
H(t_2)a(t_2) = H(t_1) a(t_1) \; .
\ee

The energy density of the long wavelength gravitational wave mode
$k_1$ during the $s_1>1$ phase is given by Eq. (\ref{rho_LW_s1}). We
write the share of the energy density in the long wavelength
gravitational wave as

\be
\label{delta_1}
\delta_1 (t) \equiv \frac{\Delta \rho_{1} (t) }{\rho (t)} 
= \epsilon(k_1) [a(t)]^{-2 + 2/s_1} \quad \quad t \leq 0 \; ,
\ee
where the constant $\epsilon (k_1)$ includes the square of the amplitude
of mode $|h(k_1)|^2$ as well as the numerical factors in 
Eq. (\ref{rho_LW_s1}). The ratio $\delta_1(t)$ is a measure of the
strength of the back reaction of long wavelength
gravity waves on the expansion rate. As previously discussed, this ratio
decays in time during inflation.

The amplitude of gravitational waves implicit in $\epsilon(k_1)$ is found
by quantizing the metric perturbations, and the well-known
result\cite{GW_Grish} is $h(k_1) \approx \kappa H(t_1)$. This is a very
small number as long as inflation happens below the Planck scale, and
even for GUT-scale inflation this amplitude is only $|h| \sim 10^{-6}$.
However, the exact value of $|h(k_1)|$ is irrelevant for our purposes: we
just assume that it is some small number.

After the transition to the decelerating phase
at $t=0$, the share of energy density in super-horizon gravity
waves is given by

\be
\label{delta_2}
\delta_2 (t) \equiv \frac{\Delta \rho_{2} (t) }{\rho (t)} 
= \epsilon(k_1) [a(t)]^{-2 + 2/s_2} \quad \quad t \geq 0 \; .
\ee
As $s_2<1$, $\delta_2$ grows with time.

Comparing (\ref{delta_1}) and (\ref{delta_2}), the shares of the total
energy density contributed by the mode $k_1$ in each of the two phases
are equal when

\be
\label{r12}
[a(t \leq 0)]^{-2+2/s_1} = [a(t \geq 0)]^{-2+2/s_2}  \; .
\ee

However, the fraction of the energy density $\delta_2(t)$ is limited from
above, since by the time
$t_2$ the gravitational wave of wavelength $\lambda^p_1$ crosses back
into the Hubble radius. At that time $t_2$ the energy density in mode
$k_1$ starts to decay as $a^{-4}$ and quickly becomes just another
radiation-like component of the energy density of the universe. From Eq.
(\ref{t_12}) we see that the time $t_2$ is defined as

\be
\label{t_12_2}
[a(t_1)]^{-1+1/s_1} = [a(t_2)]^{-1+1/s_2} \; .
\ee
Therefore, from Eq. (\ref{r12}) the (rather small) fraction of energy
density $\delta_1(t_1)$ when the scale $k_1$ crossed the Hubble radius
during inflation is equal to the maximal fraction of energy density due
to that gravitational wave during the decelerating phase, $\delta_2^{\rm
max}=\delta_2(t_2)$.

In summary, we found that when the universe inflates and then reheats,
the energy density in gravitational waves goes through 4 periods: first,
when the gravitational wave is still well inside the Hubble radius during
inflation, its energy density is essentially kinetic and falls like
radiation, $\Delta \rho \propto a^{-4}$.  Second, after the wave crosses
the Hubble radius, its energy density falls like $a^{-2}$ and is
negative, since the main contribution comes now from the drag term
$H\dot{h}h$. The same behavior $\Delta \rho \propto a^{-2}$ persists in
the third phase, when the universe reheats and starts to expand at a
decelerating rate. Therefore, after inflation the share of total energy
density due to long wavelength gravitational waves increases. Finally, by
the time that the share of the energy density in gravitational waves
approaches the value of the share of the energy density when the mode
first crossed the Hubble radius during inflation $H^{-1}(t_1)$, the mode
crosses back into the Hubble radius $H^{-1}(t_2)$ and starts to behave
like an ordinary radiation component.


\section{Conclusions}

We have discussed the energy density in gravitational waves, both short
and long wavelength. We found that ignoring the energy density and
pressure of long wavelength gravitational waves is tantamount to
violating energy conservation. We have also argued that these
interactions are not in profanation of causality or locality: on the
contrary, the persistence of gravitational interactions after the
wavelength of a perturbation becomes larger than the Hubble radius is
mandated by time-reversal invariance of the classical equations.

The energy density and pressure of long wavelength gravitational waves
tend to slow the expansion rate of the universe, and their share of the
total energy density and pressure grows in time during periods of
decelerated expansion ($s<1$). However, this share has an upper limit
during the decelerated expansion phase which is the value of that share
at the instant when the gravitational wave mode crossed the Hubble radius
for the first time during inflation. As soon as this limit is reached,
the mode crosses back into the Hubble radius, its energy density begins
to fall like $a^{-4}$ and the gravitational wave starts to behave like
ordinary radiation. 

Our results are in many ways similar to those of Sahni\cite{Sahni}, where
the energy density of gravitational waves during a decelerated expansion
phase is computed but only insofar as the gravitational waves are inside
the Hubble radius. We have shown that this implies non-conservation of
energy, and we indicated that the most natural way of curing this
pathology is simply to include the energy density of gravitational waves
irrespective of their wavelength.

\vskip 0.5cm 
\noindent {\bf Acknowledgements}

I would like to thank R. Brandenberger, S. Mukhanov, R. Woodard and M.
Maia for stimulating conversations. This work was supported by the U.S.
DOE under Contract DE-FG0297ER41029, Task A at the University of Florida.



\begin{thebibliography}{99}

\bibitem{Sahni} V. Sahni, {\it Phys. Rev.} {\bf D42}, 453 (1990).

\bibitem{GW_Grish} L. Grishchuk, {\it Zh. Eksp. Teor. Fiz.}
{\bf 67}, 825 (1974) [{\it Sov. Phys. JETP} {\bf 40}, 409 (1975)];
{\it Phys. Rev.} {\bf D48}, 5581 (1993).

\bibitem{GW_Starob} A. Starobinsky, {\it Pis'ma Zh. Eksp. Teor. Fiz.}
{\bf 30}, 719 (1979) [{\it JETP Lett.} {\bf 30}, 682 (1979)].

\bibitem{GW_Inf1} L. Abbott and D. Harari, {\it Nucl. Phys.} {\bf B264},
487 (1986).

\bibitem{GW_Inf2} B. Allen, {\it Phys. Rev.} {\bf D37}, 2078 (1988).

\bibitem{GW_CMB} V. Rubakov, M. Sazhin and A. Veryaskin,
{\it Phys. Lett.} {\bf B115}, 189 (1982); R. Fabbri and M. Pollock,
{\it ibid.} {\bf B125}, 445 (1983); L. Abbott and M. Wise, {\it
ibid.} {\bf B135}, 279 (1984).

\bibitem{GW_LIGO} A. Abramovici {\it et al.} , {\it Science} {\bf 256},
325 (1992).

\bibitem{TW} N. Tsamis and R. Woodard, {\it Phys. Lett.} {\bf B301}, 351
(1993); N. Tsamis and R. Woodard, {\it Nucl. Phys.} {\bf B474}, 235
(1996). 

\bibitem{MAB}  V. Mukhanov, L.R. Abramo and R. Brandenberger, {\it
Phys. Rev. Lett.} {\bf 78}, 1624 (1997).

\bibitem{ABM} L.R. Abramo, R. Brandenberger and V.
Mukhanov, {\it Phys. Rev.} {\bf D56}, 3248 (1997).

\bibitem{AW} L. R. Abramo and R. Woodard, preprints astro-ph/9811430
and astro-ph/9811431 (1998).

\bibitem{Weinberg} S. Weinberg, ``Gravitation and Cosmology" (Wiley, New
York 1972).

\bibitem{revpaper}  V. Mukhanov, H. Feldman and R. Brandenberger, {\it
Phys. Rep.} {\bf 215}, (1992), 203-333.

\bibitem{Isaacson} R. Isaacson, {\it Phys. Rev.} {\bf 166}, 1263 (1968);
{\bf 166} 1272 (1968).

\bibitem{ITTW} J. Iliopoulos, T. Tomaras, N. Tsamis and R. Woodard, {\it
Nucl. Phys.} {\bf B534}, 419 (1998).

\bibitem{Maia} M. Maia and J. Barrow, {\it Phys. Rev.} 
{\bf D50}, 6262 (1994); M. Maia, J. Carvalho and J. Alcaniz, 
{\it ibid} {\bf D56}, 6351 (1997); M. Maia, J. Carvalho and J.
Alcaniz, preprint astro-ph/9902338.

\end{thebibliography}
\end{document}